\documentclass[letterpaper]{article} 
\usepackage{aaai2026}  
\usepackage{times}  
\usepackage{helvet}  
\usepackage{courier}  
\usepackage[hyphens]{url}  
\usepackage{graphicx} 
\usepackage{natbib}  
\usepackage{caption} 
\usepackage{algorithm}
\usepackage{algorithmic}

\usepackage{newfloat}
\usepackage{listings}
\DeclareCaptionStyle{ruled}{labelfont=normalfont,labelsep=colon,strut=off} 
\floatstyle{ruled}
\newfloat{listing}{tb}{lst}{}
\floatname{listing}{Listing}
\usepackage{xcolor}
\newcommand{\answerYes}[1]{\textcolor{blue}{#1}} 
\newcommand{\answerNo}[1]{\textcolor{teal}{#1}} 
\newcommand{\answerNA}[1]{\textcolor{gray}{#1}}

\title{Community Norms in the Spotlight: Enabling Task-Agnostic Unsupervised Pre-Training to Benefit Online Social Media}
\author{
    Liam Hebert,
    Lucas Kopp,
    Robin Cohen
}
\affiliations{
    Cheriton School of Computer Science\\
    University of Waterloo \\
    Waterloo, Ontario, Canada \\
    liam.hebert@uwaterloo.ca

}

\usepackage{bibentry}

\begin{document}

\maketitle

\begin{abstract}
Modelling the complex dynamics of online social platforms is critical for addressing challenges such as hate speech and misinformation. While Discussion Transformers, which model conversations as graph structures, have emerged as a promising architecture, their potential is severely constrained by reliance on high-quality, human-labelled datasets. In this paper, we advocate a paradigm shift from task-specific fine-tuning to unsupervised pretraining, grounded in an entirely novel consideration of community norms. We posit that this framework not only mitigates data scarcity but also enables interpretation of the social norms underlying the decisions made by such an AI system. Ultimately, we believe that this direction offers many opportunities for AI for Social Good.
\end{abstract}


\section{Introduction}
Online social platforms have become central arenas for human interaction, fostering everything from collaborative problem-solving to widespread public discourse. Modelling the complex dynamics of these interactions is a fundamental challenge to hate speech, especially in hate speech detection, where context can dramatically alter meaning. To move beyond analyzing isolated comments, models must understand the context in which a statement is made. Notably, we argue that useful context is multifaceted, including not only semantic meaning and structure of entire conversations, but also the subtle, implicit social and community norms that govern behaviour. Discussion Transformers, which model the hierarchical, branching structure of conversations using Graph Transformers, have emerged as a promising architecture for capturing this rich contextual information \citep{hebert2024multi, hebert2022predicting}.

Despite their structural advantages, a significant bottleneck limits the potential of current Discussion Transformers: their heavy reliance on supervised fine-tuning to learn how to interpret context. The availability of high-quality, human-labelled data inherently constrains this approach. Notably, existing labelled datasets that focus on individual comments cannot be used with Discussion Transformers, as these models require complete discussion trees for training, including multimodal information. For complex social phenomena, such as hate speech or misinformation, such data sets are notoriously sparse and creating them is time consuming and expensive. All this is exacerbated by an intrinsic ``cat-and-mouse game'' in which platforms are actively incentivized to remove the very content needed for training, further shrinking available datasets. For example, the Contextual Abuse Dataset (CAD) was proposed in 2021, comprising 1394 discussions and 27,487 labels, and now only consists of 183 discussions and 2,765 labels (a $\approx$ 10x reduction) \cite{vidgen-etal-2021-introducing} \footnote{Analysis conducted using the official Reddit API in Nov 2025}. 

Concurrently, we also argue that this fine-tuning paradigm is inherently limited for learning the full spectrum of social community norms. Datasets for fine-tuning Discussion Transformers are often narrowly focused on accomplishing a single task (e.g. hate speech detection) \cite{hebert2024multi}. They may lack representative diversity in online communities, inequitably covering the political spectrum or geographic regions, thereby hindering the model's robustness and its potential to understand the nuanced social dynamics intended to be captured.

In this paper, we advocate a paradigm shift: moving the heavy lifting of contextual understanding from task-specific fine-tuning to task-agnostic, unsupervised pre-training. Akin to how humans learn to walk before we run, we propose that Discussion Transformers should first understand the foundational principles of conversational context, from linguistic semantics to interaction patterns, by leveraging the vast, unlabeled corpus of online discussions. This approach mirrors the success of pre-training in Natural Language Processing (NLP), which has proven that learning transferable skills from unlabeled data is a powerful solution to data scarcity \citep{radford2021learning}. In particular, we argue that pre-training tasks can be framed through the lens of community norm comprehension (e.g. \citet{griffiths2010norm}), training models that understand not just \textit{what} is being said, but \textit{how} and \textit{why} it is said within a specific social norm.

To realize this vision, we build a conceptual framework in which pre-training tasks are explicitly grounded in learning community norms. We bridge established principles from generative and contrastive pre-training in both Natural Language Processing (NLP) \cite{radford2021learning} and Graph Neural Networks (GNNs) \cite{hu2020gpt, you2020graph}, re-adapting them to the unique topological and social properties of discussion data. We theorize how novel generative tasks (e.g., edge-level reply classification) can teach a model the structural norms of conversation. In contrast, discussion-aware contrastive tasks (e.g., sampling and comparing distinct discussion branches from the same root) can enable learning of global, high-level values and norms of diverse communities. We argue that this framework, by combining local (generative) and global (contrastive) norm learning, enables models to develop the complementary skills needed to comprehend social discussions. We also discuss the significant risk of negative transfer --- a known pitfall in GNN pre-training \cite{donabauer2024challenges} --- and argue that it often stems from tasks misaligned with the desired social-level understanding. We thus discuss the critical need to design these tasks to be ``discussion-aware'' to ensure positive, generalizable learning.

Finally, we argue that this norm-centric pre-training framework not only enables more robust, data-efficient models but also opens new, critical avenues for analysis. We conclude by discussing how this approach, when coupled with explainable AI (XAI) techniques, could enable us to interpret the social norms these models learn directly. This interpretability is a crucial step toward auditing models for bias and fostering the development of autonomous systems that promote equitable, safe, and positive social good in online communities \citep{fisher2021towards}.

\section{A Framework for Norm-Centric Pre-Training}
\label{sec:framework}
Instead of relying on task-specific labels to indirectly teach models about social dynamics, we advocate a framework in which the primary, explicit goal of pre-training is to learn transferable, foundational models of community norms.
Our core ``blue sky'' hypothesis is that by leveraging vast amounts of unlabeled discussion data, we can train models to first understand the social norms of online communities before using task-specific, potentially biased, labelled data.

Our proposed framework is directly inspired by the success of pre-training in Natural Language Processing (NLP) and Graph Neural Networks (GNNs), which have demonstrated their potential to address data scarcity. Pre-training tasks can broadly be categorized as either generative or contrastive. On one hand, \textbf{generative pre-training}, such as Masked Language Modelling (MLM) \cite{devlin2019bert} or node feature reconstruction \cite{hu2020gpt}, trains models to understand local structure and semantics by reconstructing masked or missing parts of the input. On the other hand, \textbf{contrastive pre-training}, such as Next Sentence Prediction (NSP) or GraphCL \cite{you2020graph}, trains models to learn global representations by pulling ``positive'' (semantically similar) pairs together and pushing ``negative'' (dissimilar) pairs apart in an embedding space. 

We argue that the duality between generative and contrastive tasks provides a powerful conceptual toolkit for pretraining discussion transformers. We can adapt generative tasks to teach Discussion Transformers the local, structural, and semantic norms of a conversation (e.g., ``What kind of reply is appropriate at this specific point in the discussion?''). Concurrently, we can design novel contrastive tasks to teach the global, high-level norms that define an entire community's unique ``fingerprint'' (e.g., ``Does this entire conversation reflect the behavioral norms of community A or community B (such as left-leaning or right-leaning)?''). In this framework, generative tasks refine the model's understanding of local conversational rules, while contrastive tasks refine its ability to capture the holistic, emergent social signatures of diverse communities. In the following subsections, we envision potential norm-aware generative and contrastive pre-training tasks for discussion transformers. 

\subsection{Generative Pre-Training for Structural and Semantic Norms}
Generative tasks excel at teaching models about local context by having them reconstruct corrupted or masked portions of the input data. In Graph Neural Networks (GNNs), for instance, generative pre-training has been successfully explored in domains such as chemistry \cite{hu2020strategies} and recommender systems. These tasks are typically defined at two levels of granularity. At the edge level, generative tasks aim to predict whether two nodes are connected in the graph, thereby emphasizing the model's grasp of the graph's structural knowledge. At the node level, tasks aim to reconstruct the representations (features) of masked nodes, thereby forcing the model to learn semantic context from a node's neighbourhood. In our norm-centric framework, we adapt the principle of edge-level and node-level generative tasks to teach a Discussion Transformer the local, micro-level rules that govern a conversation. These include both \textbf{structural norms} (i.e., ``How do replies connect?'') and \textbf{semantic norms} (i.e., ``What kind of content is expected at this point in the discussion?'').

To learn structural norms, we propose an edge-level ``conversation structure'' classification task. In its simplest form, this task would adapt GNN-style edge masking by removing directed reply-to edges from the discussion graph and tasking the model with classifying whether two comments are intended to be connected (i.e., whether one is a valid reply to the other). We hypothesize that this task would directly stress the model's ability to learn the structural conventions within a community. While this ``is a reply'' task might seem trivial for a traditional Graph Neural Network (GNN) --- whose message-passing mechanism inherently ``knows'' the graph structure \citep{velickovic2018graph, hamilton2017inductive, chen2018fastgcn} --- it is a non-trivial and powerful learning signal for a Graph Transformer \cite{ying2021transformers}. Because Transformers rely on learned positional encodings to understand structure, this task forces the model to imbue those encodings with rich, meaningful information about the discussion's topology. 

Next, to learn community semantics, we propose a node-level ``semantic norm'' reconstruction task. Analogous to Masked Language Modelling (MLM) in text \cite{devlin2019bert} or node attribute reconstruction in GNNs \cite{hu2020gpt}, this task would involve masking an entire comment's node features (i.e., its text/multimodal embedding) and requiring the model to reconstruct it using the context of the surrounding discussion. This objective directly models the local semantic norms of a conversation, forcing the model to predict the kind of content that is ``supposed'' to appear in the masked comment's place. We theorize that this task can help the model learn a community's social norms, imbuing it with the kinds of replies appropriate to a given conversational context.

Both proposed tasks present significant ``blue sky'' research challenges specific to their applications in complex social discussions. First, on the edge-level ``is a reply'' classification, it is key to recognize that there are often multiple plausible replies to a given comment. For instance, the reply ``I agree'' could apply to any comment expressing an opinion, even if the ground-truth conversation contains a different reply. This ambiguity suggests that future work should explore flexible loss functions that can accommodate multiple correct answers, rather than a single, rigid ground truth.

The node-level ``comment reconstruction'' task also presents a challenge when applied to social networks, stemming from its typical use in molecular modelling. This is primarily due to the fundamental difference in information density between text and an atom. As noted in molecular GNN pre-training \cite{hu2020strategies}, an atom's features are sparse and constrained (e.g., atomic number, chirality). A comment, by contrast, is composed of unbounded language (and potentially images), carrying virtually limitless semantic possibilities. We hypothesize that reconstructing such a dense, high-information representation is orders of magnitude more difficult than reconstructing a sparse atom embedding. However, a model that succeeds at this task would possess a deep, generative understanding of local conversational context. The development of robust transformer architectures and loss functions capable of achieving this feat would be a step forward.

\subsection{Contrastive Pre-Training for Global Community Norms}
\label{sec:contrastive}
While generative tasks refine a model's understanding of local rules, contrastive tasks are designed to teach it about global, high-level representations. They train the model to distinguish between inputs that are semantically similar (positive pairs) and those that are dissimilar (negative pairs). In our framework, we leverage this mechanism to move beyond local semantics and teach the Discussion Transformer the global, emergent social norms that define a discussion or even a community.

First, we propose a ``Discourse Norms'' Branch Sampling task that adapts the concept of ``subgraph sampling'' from GNN pre-training (e.g., GraphCL \cite{you2020graph}) to the unique tree structure of discussions. To create a positive pair, we would first sample two distinct branches stemming from the same root post. These two branches represent distinct sub-conversations that, while topically distinct, are both governed by the same initial context and ``discourse norms'' of the parent thread. To prevent the model from ``cheating'' by simply matching the identical root post, the root itself could be masked or removed from each sampled branch. The model is then trained using a contrastive loss, such as InfoNCE \cite{oord2018representation}, to pull the representations of these two related branches closer together while simultaneously pushing them away from other branches (negatives) sampled from entirely different discussions in the batch. Success at this task requires the model to learn a global representation that captures the internal coherence and shared context of a single conversation.

Second, we propose a high-level ``Community Norms'' Alignment task. While the previous task operates within a single discussion, this task operates across discussions to explicitly learn the unique ``social fingerprint'' of a community. We explicitly learn this by formulating a contrastive task in which the community of origin defines the positive pairs. Specifically, the model is presented with two different discussion trees sampled from the same community (e.g., two distinct threads from r/science) as a positive pair, and two discussion trees sampled from two different communities (e.g., one from r/science and one from r/politics) as a negative pair. By training the model to pull positive embeddings together, optimizing an InfoNCE loss \cite{oord2018representation}, we can force the global graph representation to encode latent properties that are invariant across the community. The model must learn to recognize the specific dialect, topics of interest, and implicit rules of engagement that distinguish one social space from another. We hypothesize that this approach can enable the model to learn a robust, comparative map of social norms, potentially reducing bias and improving performance on downstream tasks such as hate speech detection, where community context is paramount.

\section{Preliminary Experiments}
\label{sec:preliminary}
\begin{figure}
    \includegraphics[width=\linewidth]{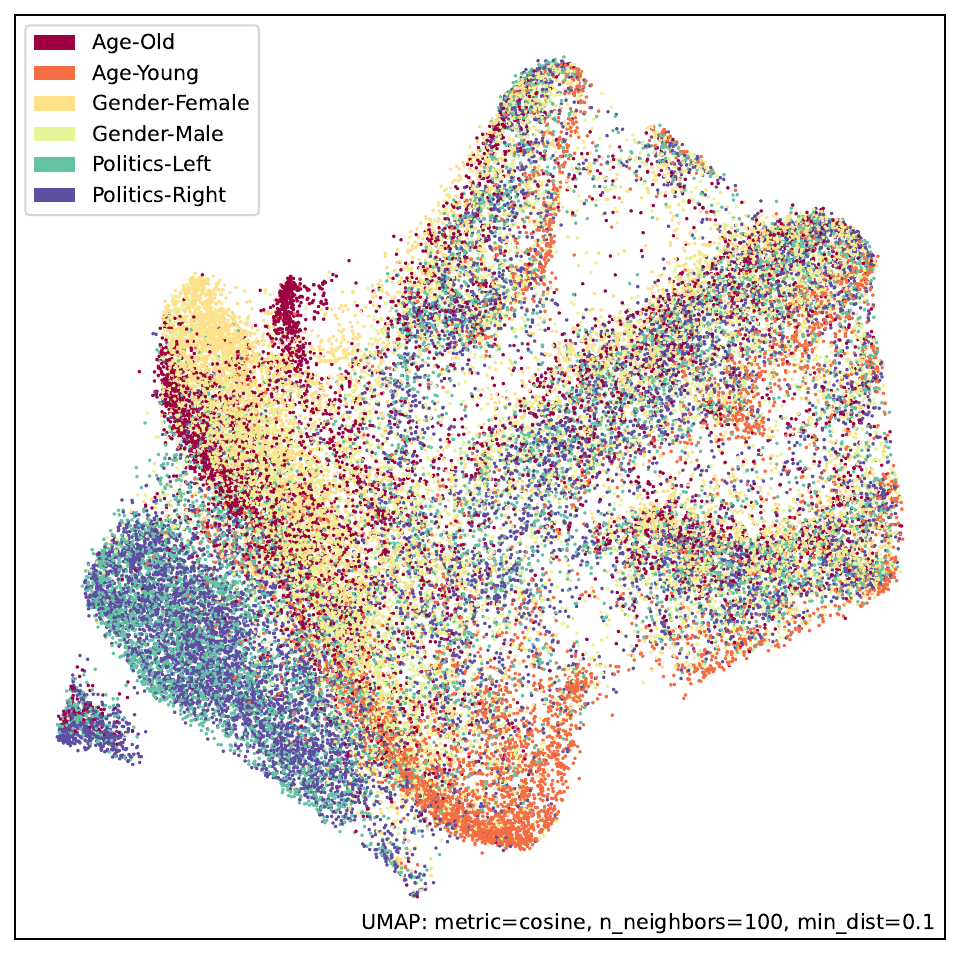}
    \caption{UMAP projection of the embedding space after applying community-centric contrastive pre-training on mDT \citep{hebert2024multi}, where each dot is a discussion. We observe a clear separation between age-oriented discussions and political similarities, suggesting the potential to capture distinct behaviours.}
    \label{fig:after_pretraining}
\end{figure}

To explore the potential of community-aware pre-training for Discussion Transformers, we conduct a preliminary study incorporating community-level contrastive pre-training on top of the
mDT Discussion Transformer 
\citep{hebert2024multi}. We formulate the pre-training task by applying a contrastive InfoNCE loss, treating discussions within the same community as positive pairs and those from other communities as negative pairs. To adapt mDT to create discussion-level embeddings, we average comment embeddings. While it would be possible to use the subreddit name directly to cluster conversations, we instead focus on broader behaviours by treating discussions from communities with high membership overlap as members of the same group. To do this, we refer to the Reddit community clusters defined by \citet{waller_quantifying_2021}, which utilize a subreddit2vec algorithm to cluster communities based on membership similarity and categorize each higher-level cluster into a specific social dimension. For our study, we focus on discussions from the ``Politics'' (left-leaning, right-leaning), ``Age'' (young, old), ``Gender'' (male, female). We sampled the top 8000 discussions from subreddits in each group, ranked by upvote score, to construct our dataset. 

Figure \ref{fig:after_pretraining} shows a visualization of the embedding space after norm-centric pre-training, where each point represents the embedding of a discussion. We can see that the model accurately clusters discussion norms, with political spheres and the age dichotomy well represented. We theorize that including generative pre-training as well would enable the model to better discern topics within the embedding space.
 
\section{Discussion and Conclusion}
\label{sec:discussion}

Our work suggests several directions for integrating explainability and interpretability techniques into a contextually aware conversation transformer. At present, conversation embeddings are created without explicit community identifiers (e.g., subreddit names), so new conversations can be assigned behavioural “prototypes” by nearest-neighbour search in embedding space. Comparing these semantically similar conversations across communities could reveal how similar behaviours are expressed. Beyond this representation-level analysis, mechanistic interpretability methods could be used to profile features of individual discussions by examining the activation patterns of specific model components (e.g., layers, attention heads, MLP neurons). By examining activated subgraphs of model components (circuits) \cite{olah_zoom_2020}, we can better understand which circuits are activated by different conversational behaviours, such as in-group jokes or insults. Moreover, dictionary learning with sparse autoencoders can decompose model activations into a sparse, approximately mono-semantic set of features that align with human-interpretable concepts \cite{huben2024sparse, bricken2023towards}. Features whose activation is strongly associated with particular communities or clusters of conversations could then be interpreted as latent ``norm dimensions'' to profile and label how members of a community tend to interact. 

The broader benefit of combining representation-level and mechanistic explanations is that they enable us to identify online behavioural prototypes and model community norms in a way that is both empirically grounded in the model’s internal activation patterns and can be inspected by humans, which philosophers have argued is important for AI safety \cite{kastner_explaining_2024}. For platform governance, this supports more context-sensitive moderation, enabling communities with norms that diverge from site-wide guidelines to be moderated at the level of conversational patterns rather than by a surface-level toxicity score, thereby minimizing the frequency with which harmless content is censored or harmful content is permitted \cite{tobi_towards_2024}. Interpretability tools could help distinguish activation patterns and feature combinations between appropriate and inappropriate uses of potentially hateful language, and profile how such language functions across different communities. For platform users, providing interpretable summaries of a community's characteristics can make the community's implicit social contract more explicit.

Since the model embeds conversations in a shared space, it can be used to study temporal behavioural and conversational dynamics, as well as community polarization. For example, prior work has shown that Reddit saw a sharp increase in political polarization around the 2016 U.S. presidential election by tracking how community embeddings and interaction networks changed over time \cite{waller_quantifying_2021}. Similar methods can be applied to conversational embeddings, enabling us to study how discussions have moved within the embedding space over time. We could determine whether clusters of political conversations have moved farther apart and whether within-cluster variance has decreased. This could provide embedding-based metrics for polarization and echo chambers, informing better prototypes of group behaviour.

It is also rewarding to identify common ground with existing research. Significant parallels arise with the vision for normative modelling of sociotechnical governance in \citet{singh2015norms}, which proposes that systems learn social states from participants' interactions. We see potential for the type of norms discussed therein to be realized through our proposed framework of contrastive and generative pre-training. We also observe efforts in NL generation \citep{hong-etal-2025-normgenesis} to synthesize culturally grounded dialogues via model norm adherence and repair; in contrast, we can leverage the vast scale of unlabelled data to learn these dynamics. We note earlier interest in the use of tags \citep{griffiths2010norm} to facilitate norm development and recognize the link to our emphasis on features to frame our valued concept of community norms. We also see connections with the idea of meta-level acceleration proposed by \citet{zhao2025towards, meta_level_zhao25}, where foundational and normative pre-training can reduce the labour intensity of analyzing complex social phenomena on social networks. In line with that AI for Social Good work, we stress the importance of representative data collection to capture diverse community norms. 

In all, we believe that future AI models that analyze social behaviour will not be defined by their ability to process text, but by their capacity to develop a deeper understanding of the social context in which that behaviour exists (i.e. seeing the trees \textbf{and} the forest). Our preliminary results demonstrate the potential to capture these norms, in ways that address data scarcity and enlighten both users and researchers about the broad landscape of social media discussions.

\section{Acknowledgements}
Hebert gratefully thanks the financial support from the Natural Sciences and Engineering Research Council of Canada (NSERC) through the Vanier Graduate Scholarship, as well as from the IEEE Nick Cercone Graduate Scholarship. The authors also thank the University of Waterloo for its institutional support.

\bibliography{aaai2026}

\pagebreak
\section{Paper Checklist}
\begin{enumerate}

\item For most authors...
\begin{enumerate}
    \item  Would answering this research question advance science without violating social contracts, such as violating privacy norms, perpetuating unfair profiling, exacerbating the socio-economic divide, or implying disrespect to societies or cultures?
    \answerYes{Implementing our proposed framework can be done on anonymized data, which is already publicly available. Our paper makes clear the argument for representative data to prevent such harms from occuring.}
  \item Do your main claims in the abstract and introduction accurately reflect the paper's contributions and scope?
    \answerYes{Yes, our framework is described in section ``A Framework for Norm-Centric Pre-Training''. Our findings and contributions are summarized in ``Discussion and Conclusion''.}
   \item Do you clarify how the proposed methodological approach is appropriate for the claims made? 
    \answerYes{Yes, our framework described in section ``A Framework for Norm-Centric Pre-Training'' is accompanied by a discussion on its applicability and forward-looking statements on the benefits it would provide.}
   \item Do you clarify what are possible artifacts in the data used, given population-specific distributions?
    \answerNo{As our paper is preliminary and late-breaking, we do not have any artifacts.}
  \item Did you describe the limitations of your work?
    \answerYes{Yes, we discuss potential limitations and challenges of implementing our framework in section ``A Framework for Norm-Centric Pre-Training''}
  \item Did you discuss any potential negative societal impacts of your work?
    \answerYes{We discuss many potential ramifications and applications of our work in section ``Discussion and Conclusion''}
      \item Did you discuss any potential misuse of your work?
    \answerYes{We discuss many potential ramifications and applications of our work in section ``Discussion and Conclusion''}
    \item Did you describe steps taken to prevent or mitigate potential negative outcomes of the research, such as data and model documentation, data anonymization, responsible release, access control, and the reproducibility of findings?
    \answerNo{As a preliminary Bluesky work, we are not releasing any data or models at this time. }
  \item Have you read the ethics review guidelines and ensured that your paper conforms to them?
    \answerYes{We acknowledge and agree to the ethics review guidelines and ensure our paper conforms to them.}
\end{enumerate}

\item Additionally, if your study involves hypotheses testing...
\begin{enumerate}
  \item Did you clearly state the assumptions underlying all theoretical results?
    \answerNA{NA}
  \item Have you provided justifications for all theoretical results?
    \answerNA{NA}
  \item Did you discuss competing hypotheses or theories that might challenge or complement your theoretical results?
    \answerNA{NA}
  \item Have you considered alternative mechanisms or explanations that might account for the same outcomes observed in your study?
    \answerNA{NA}
  \item Did you address potential biases or limitations in your theoretical framework?
    \answerNA{NA}
  \item Have you related your theoretical results to the existing literature in social science?
    \answerNA{NA}
  \item Did you discuss the implications of your theoretical results for policy, practice, or further research in the social science domain?
    \answerNA{NA}
\end{enumerate}

\item Additionally, if you are including theoretical proofs...
\begin{enumerate}
  \item Did you state the full set of assumptions of all theoretical results?
    \answerNA{NA}
	\item Did you include complete proofs of all theoretical results?
    \answerNA{NA}
\end{enumerate}

\item Additionally, if you ran machine learning experiments...
\begin{enumerate}
  \item Did you include the code, data, and instructions needed to reproduce the main experimental results (either in the supplemental material or as a URL)?
    \answerNo{As our work is preliminary and exploratory, we do not have any code or data to open-source at this time. We plan to continue this line of work and will release the code and data upon completion.}
  \item Did you specify all the training details (e.g., data splits, hyperparameters, how they were chosen)?
    \answerNA{We do not conduct exhaustive experiments and only report a preliminary result, fitting the theme of a late-breaking exploratory paper, depicted in Figure \ref{fig:after_pretraining}, discussed in section ``Preliminary Experiments''.}
     \item Did you report error bars (e.g., with respect to the random seed after running experiments multiple times)?
    \answerNA{We do not conduct exhaustive experiments and only report a preliminary result, fitting the theme of a late-breaking exploratory paper.}
	\item Did you include the total amount of compute and the type of resources used (e.g., type of GPUs, internal cluster, or cloud provider)?
    \answerNo{We do not conduct exhaustive experiments and only report a preliminary result, fitting the theme of a late-breaking exploratory paper. However, we note that we used 1 Ada 6000 GPU on an internal cluster for our experiments.}
     \item Do you justify how the proposed evaluation is sufficient and appropriate to the claims made? 
    \answerNA{Our results are preliminary and serve to motivate future work on the ideas presented in our proposed framework.}
     \item Do you discuss what is ``the cost`` of misclassification and fault (in)tolerance?
    \answerYes{We discuss many implications of our work in section ``Discussion and Conclusion''.}
  
\end{enumerate}

\item Additionally, if you are using existing assets (e.g., code, data, models) or curating/releasing new assets, \textbf{without compromising anonymity}...
\begin{enumerate}
  \item If your work uses existing assets, did you cite the creators?
    \answerYes{In section ``Preliminary experiments'', we use the social clusters provided by \cite{waller_quantifying_2021}, and cite them accordingly.}
  \item Did you mention the license of the assets?
    \answerNA{There is no license attached to \cite{waller_quantifying_2021}.}
  \item Did you include any new assets in the supplemental material or as a URL?
    \answerNA{We do not release any new datasets or assets.}
  \item Did you discuss whether and how consent was obtained from people whose data you're using/curating?
    \answerNA{We do not release any new datasets.}
  \item Did you discuss whether the data you are using/curating contains personally identifiable information or offensive content?
    \answerNA{We do not release any new datasets.}
\item If you are curating or releasing new datasets, did you discuss how you intend to make your datasets FAIR?
\answerNA{We do not release any new datasets.}
\item If you are curating or releasing new datasets, did you create a Datasheet for the Dataset? 
\answerNA{We do not release any new datasets.}
\end{enumerate}

\item Additionally, if you used crowdsourcing or conducted research with human subjects, \textbf{without compromising anonymity}...
\begin{enumerate}
  \item Did you include the full text of instructions given to participants and screenshots?
    \answerNA{We did not use crowdsourcing or research with human subjects.}
  \item Did you describe any potential participant risks, with mentions of Institutional Review Board (IRB) approvals?
    \answerNA{We did not use crowdsourcing or research with human subjects.}
  \item Did you include the estimated hourly wage paid to participants and the total amount spent on participant compensation?
    \answerNA{We did not use crowdsourcing or research with human subjects.}
   \item Did you discuss how data is stored, shared, and deidentified?
   \answerNA{We did not use crowdsourcing or research with human subjects.}
\end{enumerate}

\end{enumerate}

\end{document}